\documentclass[12pt,preprint]{aastex}

\newcommand{\gsim}{\mbox{\hspace{.2em}\raisebox{.5ex}{$>$}\hspace{-.8em}\raisebox{-.5ex}{$\sim$}\hspace{.2em}}}
\newcommand{\ssst}{\scriptscriptstyle}
\newcommand{\E}[1]{\times 10^{#1}}
\newcommand{\lt}{\left} \newcommand{\rt}{\right}

  \newcommand{\ps}{\,{\rm s}^{-1}}
\newcommand{\yr}{\,{\rm yr}}    
\newcommand{\cm}{\,{\rm cm}}    \newcommand{\km}{\,{\rm km}}

\newcommand{\parsec}{\,{\rm pc}}\newcommand{\kpc}{\,{\rm kpc}}
\newcommand{\K}{\,{\rm K}}      \newcommand{\ergs}{\,{\rm ergs}}
\newcommand{\keV}{\,{\rm keV}}

\newcommand{\den}{\rho_{\ssst 0}}  
\newcommand{\no}{n_{\ssst 0}}	\newcommand{\denej}{\bar{\rho}_{\rm ej}}
\newcommand{\rj}{r_{\rm j}}	\newcommand{\tj}{t_{\rm j}}

\newcommand{\Ms}{M_{\rm s}}             \newcommand{\Ts}{T_{\rm s}}
\newcommand{\rs}{r_{\rm s}}             \newcommand{\vs}{v_{\rm s}}
\newcommand{\mH}{m_{\ssst\rm H}}	\newcommand{\Tc}{T_{\rm c}}
\newcommand{\Eo}{E_{\ssst 0}}\newcommand{\ETc}{E_{\ssst 0}}
\newcommand{\rc}{r_{\rm c}}	\newcommand{\tc}{t_{\rm c}}
\newcommand{\fAv}{F_{v}^{\rm\ssst (A)}} \newcommand{\fAr}{F_{r}^{\rm\ssst (A)}}
\newcommand{\fRv}{F_{v}^{\rm\ssst (R)}} \newcommand{\fRr}{F_{r}^{\rm\ssst (R)}}
	
\newcommand{\ASCA}{{\sl ASCA}}	
\newcommand{\lambdj}{\lambda_{\rm j}} \newcommand{\lambdc}{\lambda_{\rm c}}
         \newcommand{\Eu}{E_{51}}

\shorttitle{SNR Crossing a Density Jump}
\shortauthors{Chen et al.}

\begin{document}

\title{SUPERNOVA REMNANT CROSSING A DENSITY JUMP: A THIN SHELL MODEL}

\author{Yang Chen\altaffilmark{1,2}, Fan Zhang\altaffilmark{1},
Rosa M.\ Williams\altaffilmark{2}, and Q.\ Daniel Wang\altaffilmark{2}}
\altaffiltext{1}{Department of Astronomy, Nanjing University, Nanjing 210093,
       P.R.China}
\altaffiltext{2}{Department of Astronomy, B619E-LGRT, 
       University of Massachusetts, Amherst, MA01003}

\begin{abstract}
The environments of supernova explosion are often inhomogeneous and there
may be jumps in their density structure. We have developed
a semi-analytic model under the thin-shell approximation for supernova
remnants that evolve crossing a density jump in the ambient medium. 
The generic evolutionary relations are presented for the blast wave
after impacting on a cavity wall, which may be produced by the
energetic stellar wind from the supernova progenitor.
The relations can also be extended to the case that the blast waves 
break out from a dense cloud if different density contrast is used. 
This model is applied to N132D, a well-known cavity-born supernova
remnant whose evolution has not yet been quantitively estimated in a
cavity scenario due to lack of model formulae, and self-consistent
physical parameters are obtained.
\end{abstract}


\section{Introduction}
The environments of supernova explosion are often inhomogeneous and there
may be jumps in their density structure.
A massive star can excavate a low density cavity with its energetic stellar
wind and ionizing radiation in the circumstellar space before it explodes.
It is also possible that supernova may explode in a dense cloud and
subsequently break out from the cloud into a low density region.
These supernova remnants (SNRs) cannot be simply treated with the canonical
evolutionary laws such as the Sedov relation (Sedov 1959).
In the former case, a blast wave collides with a cavity wall after it
expands effortlessly into the cavity.
A similar situation has been suggested to have occurred in various supernova
remnants such as N132D (Hughes 1987).
In the latter case, the blast wave propagates at a higher velocity
after it breaks out of the cloud, as is suggested for 3C~391
(Reynolds \& Moffett 1993).
It would be of interest to develop an analytic model showing how
the evolution of such kind of supernova remnants deviates from the canonical
solution.

In a study of SNR evolution in a dense molecular cloud, Wheeler et al.\
(1980) briefly discussed the effect of a pre-SN cavity, suggesting that the
remnant will evolve rapidly into the radiative phase after it encounters
the cavity wall.
Chevalier and Liang (1989) presented a self-similar model for 
both the structure of the shocked powerlaw ejecta and
the shock propagation into the circumstellar shell.
However, the ejecta structure becomes unimportant when the mass swept up
by the blast wave significantly exceeds the mass of the ejecta.
Investigation of supernova blast waves crossing a density jump
has been carried out by hydrodynamic simulation.
Tenorio-Tagle et al.\ (1990, 1991) made numerical simulations of the
supernova shock interacting with a wind-driven shell (see also a review by
Franco et al.\ 1991 and references therein).
Tenorio-Tagle, Bodenheimer, \& Yorke (1985) have also presented a
numerical simulation of the evolution of a remnant resulting from
supernova explosions in or near molecular clouds.

An analytical model would not only complement the numerical
simulation, but would allow for exploring a large parameter space and
more directly provide physical insights into the evolution of this
sort of SNRs.
It has been shown by many authors (e.g., McCray 1987,
Blinnikov, Imshennik, \& Utrobin 1982, and others) that a thin-shell model
(Kompaneets 1960), though approximate, is convenient and valid
in revealing the general evolution of the supernova blast wave.
The approach was applied in a numerical algorithm to
calculate the evolution of SNRs in a density gradient
(see Wang et al.\ 1992).
This method has also been applied analytically to other explosive events,
e.g., a cosmic ray blast wave (Morfill \& Drury 1981).
Therefore we use the same approach to obtain an analytic/semi-analytic
approximation to seek the basic evolutionary rule of the blast
wave after an impact on a cavity wall or a breakout from a cloud.

\section{Thin-shell Model of Impact on a Wall}

For the ambient environment of supernova, we assume a density structure of
medium $\rho=\rho_{i}=1.4n_{i}\mH$ for $r<\rj$ and $\rho=\den=1.4\no\mH$ for
$r\ge\rj$ (where $n_{i}$ and $\no$ are the number densities of H atoms
inside and outside the density jump, respectively).
In this scenario, the ejecta or blast wave is assumed to expand easily until it
hits upon the cavity wall at $\rj$.
By the time the ejecta or blast wave arrives at $\rj$, the SNR could be in
the Sedov phase or even still in the free expansion phase, depending
on the parameters $\rj$ and $\rho_{i}$.
Denote the density ratio by $\beta\equiv(\rho_{i}+\denej)/\den$, where $\denej$
is the average density of the ejecta mass over the cavity within $\rj$,
so here we discuss the case $\beta<1$.
The swept-up material is assumed to be in a thin shell and the mass
in a conical section of solid angle $\Delta\Omega$ is given by
\begin{equation}\label{eq:ma}
\Delta\Ms=\frac{1}{3}\Delta\Omega\den\lt[\rs^{3}-\rj^{3}(1-\beta)\rt].
\end{equation}
On the assumption that most of the explosion energy is thermalized
by the time when the blastwave impacts the wall or soon after that,
the thermal pressure in a spherical remnant is given by
\begin{equation}\label{eq:pr}
p=\frac{(\gamma-1)E}{4\pi\rs^{3}/3}=\frac{E}{2\pi\rs^{3}},
\end{equation}
where $E$ is the total energy and the adiabatic index $\gamma=5/3$.
The momentum equation for the conical section is
\begin{equation}\label{eq:mo}
\frac{d}{dt}(\Delta\Ms\vs)=p\rs^{2}\Delta\Omega,
\end{equation}

From the equations (\ref{eq:ma}), (\ref{eq:pr}), and (\ref{eq:mo}), we have
\begin{equation}\label{eq:odeA}
\rs\frac{d}{dt}\lt[\lt(\rs^{3}-\rj^{3}(1-\beta)\rt)\frac{d\rs}{dt}\rt]
=\frac{3E}{2\pi\den}.
\end{equation}

\subsection{Adiabatic phase}
Before the radiative phase, the supernova remnant is assumed to be
adiabatic and the energy to remain unchanged: $E\approx\Eo$.
Then eq.(\ref{eq:odeA}) has a solution
\begin{equation}\label{eq:ast}
\vs^{2}=\lt(\frac{d\rs}{dt}\rt)^{2}=\lt(\frac{\Eo}{\pi\den}\rt)
\frac{\rs^{3}-3\rj^{3}(1-\beta)\ln\rs+C}{\lt[\rs^{3}-\rj^{3}(1-\beta)\rt]^{2}},
\end{equation}
where integral constant $C$ must be determined according to the initial
condition at $\rs=\rj$. The problem can be divided into two cases
presented below.

In the first case, the supernova remnant has already been in the Sedov phase
by the time $\tj$ when $\rs=\rj$.
The Sedov evolution law
under the thin shell approximation is given by
\(\rs=(\xi\Eo/\rho_{i})^{1/5}t^{2/5}\) where $\xi=25/4\pi$ (e.g.\ McCray 1987),
so the initial velocity at $\rj$ is $\vs^{2}=\Eo/(\pi\rho_{i}\rj^{3})$.
The blast wave velocity is thus
\begin{equation}\label{eq:vAr1}
\frac{d\rs}{dt}=\sqrt{\frac{\Eo}{\pi\den}}\frac{\sqrt{\rs^{3}-\rj^{3}(1-\beta)
[1-3\ln(\rj/\rs)]}}{\rs^{3}-\rj^{3}(1-\beta)}
\end{equation}
namely
\begin{equation}\label{eq:vAr2}
\vs=\lt(\frac{\Eo}{\pi\den\rs^{3}}\rt)^{1/2}\fAv(\lambdj),
\end{equation}
where $\lambdj\equiv\rj/\rs$ and
\begin{equation}\label{eq:fAv1}
\fAv(\lambdj)\equiv\frac{\sqrt{1-\lambdj^{3}(1-\beta)(1-3\ln\lambdj)}}
{1-\lambdj^{3}(1-\beta)}.
\end{equation}

In the second case, the remnant is still in the free expansion stage
when arriving at $\rj$.
In this case, $\rho_{i}<\denej$;
and, for simplicity, $\beta\approx0$ is assumed.
For $\beta=0$, the above initial condition cannot be used, and instead,
the convergence of eq.(\ref{eq:ast}) at $\rs=\rj$ is entailed, so
$C=-\rj^{3}+3\rj^{3}\ln\rj$.
Thus, interestingly, the velocity is still described by
eqs.(\ref{eq:vAr1})--(\ref{eq:fAv1})
(but with $\beta=0$).

Eq.(\ref{eq:vAr1}) gives rise to
\begin{equation}\label{eq:tAr}
t-\tj = \sqrt{\frac{\pi\den}{\Eo}}
\int^{\rs}_{\rj}\frac{r^{3}-\rj^{3}(1-\beta)}{\sqrt{r^{3}-\rj^{3}(1-\beta)
[1-3\ln(\rj/r)]}}\,dr
\end{equation}
or
\begin{equation}\label{eq:rAt}
\lt[\fAr(\lambdj)\rt]^{2/5}\rs
=\lt(\frac{\Eo}{\pi\den}\rt)^{1/5}(t-\tj)^{2/5},
\end{equation}
where
\begin{equation}\label{eq:fAr1}
\fAr(\lambdj)\equiv\int^{1}_{\lambdj}\frac{\lambda^{3}-\lambdj^{3}(1-\beta)}
{\sqrt{\lambda^{3}-\lambdj^{3}(1-\beta)[1-3\ln(\lambdj/\lambda)]}}\,d\lambda
=\int_{\lambdj}^{1}\frac{\lambda^{3/2}}{\fAv(\lambdj/\lambda)}\,d\lambda
\end{equation}
with $\lambda=r/\rs$.
The above integration (at $\rj\neq0$) can only be performed numerically,
as plotted in Fig.1.

Note that, for the special case $\rj=\lambdj=\tj=0$,
$\fAv=1$ and $\fAr=2/5$,
so we immediately come from eq.(\ref{eq:rAt}) back to the Sedov solution 
\(\rs=(\xi\Eo/\den)^{1/5}t^{2/5}\) where $\xi=25/4\pi$ under the thin-shell
approximation (e.g.\ McCray 1987).

As shown in Fig.1, both functions $\fAv(\lambdj)$ and $\fAr(\lambdj)$
vary with parameter $\beta$.
For $\beta\neq0$, function $\fAv(\lambdj)$ starts from $\beta^{-1/2}$,
as a limit value found for eq.(\ref{eq:fAv1})
at $\lambdj=1$ or $\rs=\rj$.
The remnant is decelerated rapidly after it enters the cavity wall, as
indicated by the ensuing rapid decrease of $\fAv(\lambdj)$.
Just when $\rs$ surpasses $\rj$,
the smaller $\beta$ is, the more drastically $\fAv$ drops
and the more quickly $\fAv$ approaches the asymptotic curve for $\beta=0$.
In addition, these asymptotic values are of order unity, implying
a crude factor of velocity decrease $\sim\beta^{-1/2}$ or a ram pressure
($\rho\vs^{2}$) roughly recovered to that before the impact
(see eq.[\ref{eq:vAr2}]).

In the limit case $\beta=0$, 
immediately after the remnant strikes the cavity wall
[$1-\lambdj\equiv(\rs-\rj)/\rs\ll1$], the behavior of $\fAv$ can be
described by the Taylor series
\begin{equation}\label{eq:fAv2}
\fAv(\lambdj)\simeq\frac{1}{\sqrt{2}}\lt[1+\frac{1}{2}(1-\lambdj)\rt],
\end{equation}
and the shock velocity at $\rs=\rj$ (i.e.\ $\lambdj=1$) is
\begin{equation}\label{eq:vw}
\vs(\rj)=\lt(\frac{\Eo}{2\pi\den\rj^{3}}\rt)^{1/2}
=1.4\E{3}\lt(\frac{\Eo}{10^{51}\ergs}\rt)^{1/2}
         \lt(\frac{\no}{1\cm^{-3}}\rt)^{-1/2}
         \lt(\frac{\rj}{5\parsec}\rt)^{-3/2}\,\km\ps,
\end{equation}
a factor $\sqrt{2}$ lower than the velocity $\vs=(\Eo/\pi\den\rj^{3})^{1/2}$
the shock would otherwise have in the Sedov solution.
Still in the case $\beta=0$, the linear relation [eq.(\ref{eq:fAv2})] 
is found to be a good approximation of $\fAv$ for $1\ge\lambdj\gsim0.2$
(see Fig.1).

At the other extremity $\lambdj\ll1$, 
$\fAv(\lambdj)$ approaches to 1 according to
\begin{equation}\label{eq:fAv3}
\fAv(\lambdj)\simeq 1+\frac{1}{2}(1-\beta)(1+3\ln\lambdj)\lambdj^{3},
\end{equation}
which allows for any values of $\beta$.

The numerical solution of function $\fAr(\lambdj)$ in eq.(\ref{eq:fAr1})
shows that it starts from 0 at $\lambdj=1$ and approaches the other extremity
$\lambdj=0$ in similar trends for various values of $\beta$ (Fig.1).
For $1-\lambdj\ll1$, in the case $\beta\neq0$,
its behavior can be approximated by a series:
\begin{equation}\label{eq:fAr4}
\fAr(\lambdj)
\approx\frac{1}{10\sqrt{\beta}}\lt[(-9+13\beta)+50(1-\beta)\lambdj
-105(1-\beta)\lambdj^{2}+4(16-17\beta)\lambdj^{5/2}\rt],
\end{equation}
which is obtained by expanding function $\fAv(\lambdj/\lambda)$
in eq.(\ref{eq:fAr1}) to the second order of the small quantity
$(1-\lambdj/\lambda)$;
in the case $\beta=0$, however,
\begin{equation}\label{eq:fAr2}
\fAr(\lambdj)\approx\frac{1}{6\sqrt{2}}\lt(3+2\lambdj
+3\lambdj^{2}-8\lambdj^{5/2}\rt).
\end{equation}
This series expansion is found valid in a broad range
$1\ge\lambdj\gsim0.3$ (Fig.1).

At the extremity $\lambdj\ll1$, function $\fAr(\lambdj)$ converges to 0.4
conforming to
\begin{equation}\label{eq:fAr3}
\fAr(\lambdj)\approx\frac{2}{5}\lt(1-\lambdj^{5/2}\rt)
             +5(1-\beta)\lt(\lambdj^{5/2}-\lambdj^{3}\rt)
             +3(1-\beta)\lambdj^{3}\ln\lambdj,
\end{equation}
which is valid for any value of $\beta$.

For $\beta\gsim0.1$, numerical integration (Fig.1) shows that $\fAr(\lambdj)$
as a whole does not vary significantly with various $\beta$.
If we take $\beta=1$ for simplicity, it corresponds to the uniform case;
eq.(\ref{eq:fAv1}) yields $\fAv=1$ and eq.(\ref{eq:fAr1}) yields
\begin{equation}\label{eq:fAr5}
\fAr(\lambdj)=\frac{2}{5}(1-\lambdj^{5/2})
\end{equation}
(see the long dashed line in Fig.1).
If this expression is adopted as a {\em crude} approximation of
$\fAr(\lambdj)$ for $\beta\gsim0.1$, eq.(\ref{eq:rAt}) then gives a
Sedov-like evolution
\begin{equation}
t-\tj\approx\sqrt{\frac{\den}{\xi\Eo}}\lt(\rs^{5/2}-\rj^{5/2}\rt),
\end{equation}
where $\xi=25/4\pi$ but can be replaced with the canonical value 2.026.

The adiabatic phase will come to an end when the gas temperature drops to
$\Tc=6\E{5}\K$ (Blinnikov et al.\ 1982). With the relation
$k\Ts=(3/16)\mu\mH\vs^{2}$ (where the mean atomic weight $\mu=0.61$),
Eq.(\ref{eq:vAr2}) with $\den=1.4\no\mH$ gives the
location $\rc$ of the phase transition:
\begin{equation}
\frac{\rc^{3}}{\lt[\fAv(\rj/\rc)\rt]^{2}}
=(4.7\parsec)^{3}\lt(\frac{\Eo}{10^{51}\ergs}\rt)
               \lt(\frac{\no}{10^{2}\cm^{-3}}\rt)^{-1}
\end{equation}
If $\beta\ll1$, the linear approximation (\ref{eq:fAv2}) is adopted and the
above formula is simplified as
\begin{equation}
\rc^{2}\rj\approx(3.8\parsec)^{3}\Eu\lt(\no/10^{2}\cm^{-3}\rt)^{-1}
\end{equation}
where $\Eu=\Eo/(10^{51}\ergs)$. 
For $\beta\ll1$, the sudden deceleration of the remnant on impacting the
cavity wall makes it possible that 
the shocked gas temperature $\Ts(\rj)$ has already gotten below $\Tc$
and the radiative phase has begun before the velocity falls
on the asymptotic curve (for $\beta\rightarrow0$, see Fig.1).
In this case, the adiabatic phase lasts so little time after the impact
that it essentially may be ignored.
Therefore it is possible that a previously freely expanding remnant
directly enters the radiative phase after it strikes the wall and
the adiabatic phase is abortive
if $\rj\gsim3.8\parsec\,\,\Eu^{1/3} (\no/10^{2}\cm^{-3})^{-1/3}$.

\subsection{Radiative phase}
When the shell is cooled down by radiation, it is assumed to be
driven by the thermal pressure of the still hot gas interior to the shell.
The internal hot gas suffers from the energy loss by the expansion:
\begin{equation}\label{eq:en}
\frac{dE}{dt}=-4\pi\rs^{2}\vs\,p.
\end{equation}
Eq.(\ref{eq:en}), inserted with eq.(\ref{eq:pr}), yields
\begin{equation}\label{eq:ET}
E=\ETc(\rc/r)^{2}.
\end{equation}
Now, eq.(\ref{eq:odeA}) becomes
\begin{equation}\label{eq:odeR}
\rs^{3}\frac{d}{dt}\lt[\lt(\rs^{3}-\rj^{3}(1-\beta)\rt)
\frac{d\rs}{dt}\rt] =\frac{3\ETc\rc^{2}}{2\pi\den},
\end{equation}
with a solution
\begin{equation}\label{eq:vRr1}
\frac{d\rs}{dt}=\sqrt{\frac{3\ETc\rc^{2}}{\pi\den}}
  \frac{\sqrt{(\rs-\rc)+\rj^{3}(\rs^{-2}-\rc^{-2})(1-\beta)/2+
  \lt[\rc-(\rj^{3}/\rc^{2})(1-\beta)
  \lt(1-3\ln(\rj/\rc)\rt)\rt]/3}}{\rs^{3}-\rj^{3}(1-\beta)}.
\end{equation}
In this solution, we have used eq.(\ref{eq:vAr1}) as the initial condition
of the shock velocity at the phase transition point $\rs=\rc$ and $t=\tc$.

Setting $\lambdc\equiv\rc/\rs$, $\eta\equiv\rc/\rj$, and $\mu=(1-\beta)/\eta^{3}$,
eq.(\ref{eq:vRr1}) is rewritten as
\begin{equation}\label{eq:vRr2}
\vs=\lt(\frac{3\ETc\rc^{2}}{\pi\den\rs^{5}}\rt)^{1/2}\fRv(\lambdc),
\end{equation}
where
\begin{equation}\label{eq:fRv1}
\fRv(\lambdc)\equiv\frac{\sqrt{(1-\lambdc)+(\mu/2)\lambdc(\lambdc^{2}-1)
 +(\lambdc/3)[1-\mu(1+3\ln\eta)]}}{1-\mu\lambdc^{3}}.
\end{equation}
Eq.(\ref{eq:vRr1}) gives
\[ t-\tc = \sqrt{\frac{\pi\den}{3\ETc\rc^{2}}}\,\,\times \]
\begin{equation}\label{eq:tRr}
 \times\int_{\rc}^{\rs}\frac{r^{3}-\rj^{3}(1-\beta)}
  {\sqrt{(r-\rc)+\rj^{3}(r^{-2}-\rc^{-2})(1-\beta)/2+
  \lt[\rc-(\rj^{3}/\rc^{2})(1-\beta)
  \lt(1-3\ln(\rj/\rc)\rt)\rt]/3}}\,dr
\end{equation}
or
\begin{equation}\label{eq:rRt}
\lt[\fRr(\lambdc)\rt]^{2/7}\rs
=\lt(\frac{3\ETc\rc^{2}}{\pi\den}\rt)^{1/7}(t-\tc)^{2/7},
\end{equation}
where
\begin{eqnarray}\nonumber
\fRr(\lambdc) &\equiv& \int_{\lambdc}^{1}\frac{\lambda^{3}-\mu\lambdc^{3}}
{\sqrt{(\lambda-\lambdc)+(\mu/2)\lambdc(\lambdc^{2}/\lambda^{2}-1)
 +(\lambdc/3)[1-\mu(1+3\ln\eta)]}}\,d\lambda\nonumber\\
 &=& \int_{\lambdc}^{1}\frac{\lambda^{5/2}}{\fRv(\lambdc/\lambda)}
   \,d\lambda.\label{eq:fRr1}
\end{eqnarray}

The numerical values of functions $\fRv(\lambdc)$ and $\fRr(\lambdc)$
are plotted in Figs.(2)--(4),
where it is seen that both functions depend on the parameters
$\beta$ and $\eta$.
For most cases of $(\eta, \beta)$ combination, $\fRv(\lambdc)$ starts from
$[1-\mu(1+3\ln\eta)]^{1/2}/[\sqrt{3}(1-\mu)]$ at $\lambdc=1$ (i.e.\
$\rs=\rc$); but for $(\eta, \beta)=(1,0)$, it starts from $6^{-1/2}$ and the
Taylor series
\begin{equation}\label{eq:fRv2}
\fRv(\lambdc)\simeq\frac{1}{\sqrt{6}}\lt[1+\frac{5}{6}(1-\lambdc)
+\frac{35}{72}(1-\lambdc)^{2}\rt]
\hspace{1cm}({\rm for}\hspace{4mm}1-\lambdc\ll1)
\end{equation}
can acts as a good approximation in the range $1>\lambdc\gsim0.5$ (Fig.2).
In the latter case, when the radiative phase begins at $\rs=\rc=\rj$
($\lambdj=\lambdc=1$), eq.(\ref{eq:vRr2}) turns back to eq.(\ref{eq:vw}).
For any $(\eta, \beta)$ combination, $\fRv(\lambdc)$
approaches 1 at the other end, $\lambdc=0$, according to
\begin{equation}\label{eq:fRv3}
\fRv(\lambdc)\simeq1-\frac{1}{12}[4+\mu(5+6\ln\eta)]\lambdc.
\end{equation}
It is noted that, at the limit $\lambdc\rightarrow0$ (or $\rs\gg\rc$),
one has $\fRv=1$ and $\fRr=2/7$, and then
eq.(\ref{eq:rRt}) is identical to the canonical solution for SNRs
in radiative phase: $\rs=(147\ETc\rc^{2}/4\pi\den)^{1/7}t^{2/7}$
(e.g.\ McCray 1987; Blinnikov et al.\ 1982).

While we could not find an analytic integration for eq.(\ref{eq:fRr1}),
we seek for a series approximation for $\lambdc\ll1$:
\[ \fRr(\lambdc) \approx \frac{1}{5040}\lt\{1440+168\lambdc(4+5\mu)
 +35\lambdc^{2}(4+5\mu)^{2}-\lambdc^{7/2}(2672+2240\mu+875\mu^{2})
              \rt.\]
\begin{equation}
 -84\lambdc\mu\lt[-12-5\lambdc(4+5\mu)+\lambdc^{5/2}(32+25\mu)\rt]\ln\eta
\lt.-1260\lambdc^{2}(-1+\lambdc^{3/2})\mu^{2}\ln^{2}\eta\rt\}\label{eq:fRr3}
\end{equation}
Eq.(\ref{eq:fRr3}) is obtained by expanding the integrand in eq.(\ref{eq:fRr1})
to the second order of small quantity $\lambdc/\lambda$ and,
for the case $\eta\gsim1.1$, is found to be very close to the numerical values
in the entire range (see Figs.3 and 4).
In the case $\eta=1$ and $\beta\neq0$,
similarly we get a series approximation for $1-\lambdc\ll1$:
\begin{equation}\label{eq:fRr4}
\fRr(\lambdc)\approx\frac{1}{70}\sqrt{\frac{3}{\beta}}
 \lt[\beta\lt(135-434\lambdc+455\lambdc^{2}-156\lambdc^{7/2}\rt)
 +7\lt(-15+54\lambdc-55\lambdc^{2}+16\lambdc^{7/2}\rt)\rt],
\end{equation}
and in the case $(\eta,\beta)=(1,0)$ ($1-\lambdc\ll1$),
\begin{equation}\label{eq:fRr2}
\fRr(\lambdc)\approx\frac{\sqrt{6}}{252}\lt(27+42\lambdc+35\lambdc^{2}\rt)
-\frac{26}{21}\sqrt{\frac{2}{3}}\lambdc^{7/2}.
\end{equation}
Eq.(\ref{eq:fRr2}) is a very good approximation for $(\eta,\beta)=(1,0)$
in the range $1\ge\lambdc\gsim0.3$ (Fig.2).



\section{Thin-shell Model of Breakout of a Cloud}
We assume a configuration of environment density similar to that
in the former case, except that $\rho_{i}>\den$ or $\beta>1$.
The material in the cloud is dense enough that we can assume the supernova
remnant has been in the adiabatic (Sedov) phase by the time of breakout.
Moreover, because the density $\den$ outside the cloud is so low, 
we need not consider the radiative phase in this scenario, as it is 
unlikely the SNR would yet have entered this phase.
Therefore most of the evolutionary solutions for the adiabatic phase
established in the former case are compatible with this case, given
a replacement of $\beta>1$.
Functions $\fAv(\lambdj)$ and $\fAr(\lambdj)$ for $\beta>1$ are plotted
in Fig.5.

\section{Applications}
\subsection{Applicability}
The model developed above can, strictly speaking, only be applied to an
SNR of spherical symmetry. In reality, an SNR is usually not symmetric,
which could be caused by the motion of the progenitor for instance. While
a good fraction of stars show a considerable proper motion, there are still
objects with a low proper motion. In practice, as long as the asymmetry is
reasonably small, our model should still provide a useful approximation.
As a general condition, the dynamical response (characterized by a sound
traveling timescale) due to the non-spherical symmetry of the density jump
surface should not be rapid enough compared with the blastwave propagation
after crossing the jump surface.

Furthermore, we have assumed that the density jump happens only once.
However, this may not be true for an SNR evolving in a stellar wind bubble,  
for example. In this case, a dense shell, surrounding a low density bubble,  
may be approximated as two density jumps. In principle, such multiple density
jumps can be incorporated into an analytic approach. But this is a bit too
complicated to be included in the present work. Here, we concentrate on
exploring the effect of one density jump on the evolution of an SNR, which
is still evolving in the shell. This particular situation is important,
because of the expected strong density jump effect on the X-ray emission
of the SNRs.

In the range of applicability discussed here, the solutions
derived in section 2 can conveniently used in estimating the
parameters of relevant supernova remnants. For example, X-ray
observations provide the X-ray luminosity or the volume emission
measure of the X-ray emitting gas as well as the gas temperature
that is related to the velocity of the blast wave.
Note that the temperature measured in a plasma model fit to the overall
spectrum of an SNR only represents a characteristic mean temperature
of the shocked gas, even for the Sedov blast wave. In principle, one
can derive a relation between this mean temperature and the shock
velocity. This problem may also be resolved to some degree with the
spatially resolved spectroscopic data, which allow for the extraction of
the spectrum of the freshly-shocked gas.
By use of eq.(\ref{eq:vAr2}) (or eq.[\ref{eq:vRr2}]) and eq.(\ref{eq:rAt})
(or eq.[\ref{eq:rRt}]) together with a relation between the volume
emission measure and the density of the ambient medium, 
any three among the five parameters, the density $\no$,
explosion energy $E$, the age of the remnant $t$, the density contrast
$\beta$, and the fractional radius of the density jump $\lambdj$,
can be derived from the other two, given the three known parameters,
the radius remnant $\rs$,
the shell velocity $\vs$, and the volume emission measure $EM$.
According to this recipes, we apply the model developed above to
SNR N132D that are conjectured to have expanded from a stellar wind
cavity into a region of dense medium and derive relevant physical
parameters.

\subsection{SNR N132D}
N132D, an SNR in the Large Magellanic Cloud, lies near the northern
boundary of an associated molecular cloud (Banas et al.\ 1997).
The inconsistency between the kinematic age and the Sedov dynamical
age led to the suggestion that the supernova explosion occurred in
a low-density cavity evacuated by the stellar wind and ionizing
radiation of a high-mass progenitor (Hughes 1987; Morse et al.\ 1996;
Hughes, Hayashi, \& Koyama 1998).
Morse et al.\ (1995) derived a kinematic age of $3150\pm200\yr$ from the
measurements of the velocities of fast moving oxygen-rich filaments.
Morse et al.\ (1996) studied the optical photoionization precursor based on
the [OIII]$\lambda5007$ surface brightness and derive a preshock density
of $\no\sim3\cm^{-3}$ and a blast shock velocity $\vs\sim800\km\ps$.
They find that these results would be trapped in unreconcilable contradictions
(either the inferred explosion energy is unusually high ($\gsim10^{52}\ergs$),
or the inferred age is much larger than the kinematic age),
unless the blast wave has been traveling much faster in the past,
probably by expanding into a low density cavity.
However, quantitive estimates of the evolution of the SNR expanding from the
cavity into the cavity wall have not yet been carried out because of
lack of proper evolutional formulae.

N132D is an oxygen-rich SNR but no evidence was found by the {\sl Hubble
Space Telescope} ({\sl HST}) for the presence oxygen-burning products.
This fact was taken by Blair et al.\ (2000) to suggest a Wolf-Rayet (WR)
star which ended at a Type Ib SN explosion. However, X-ray data from
{\sl XMM-Newton} (Behar et al.\ 2001; Aschenbach 2002) do show the
presence of significant proportions of Si and Fe. If these elements 
are actually present in the ejecta, rather than the surrounding ISM,
this might tend to contradict the scenario of a Type Ib explosion.
Earlier X-ray findings by Hwang et al. (1993) using the Focal Plane
Crystal Spectrometer on {\sl Einstein} also found strong Fe lines,
which the authors used to support the scenario of a massive progenitor
(with masses over $20M_{\odot}$ favored).
Therefore we assume that the core collapse SN explosion occurred in a
stellar wind cavity blown by a massive progenitor, possibly a WR star.

The recent {\sl XMM-Newton} X-ray observation of N132D shows a clear shell
feature from which most of the ion emission lines arise and a centrally
confined Fe~K emission charateristic of a high temperature component
(Behar et al.\ 2001; Aschenbach 2002). According to the \ASCA\ data 
analysis (Hughes et al.\ 1998), the high X-ray luminosity 
($\sim3\E{37}\ergs\ps$) of the shell comes from
the shock interaction with the dense material at the cavity wall,
with an X-ray spectral normalization factor
$N_{s}\sim5\E{12}\cm^{-5}$, and hence
a volume emission measure $EM=4\pi d^{2}N_{s}\sim1.5\E{60}\cm^{-3}$
(where $d\approx50\kpc$) (Hughes et al.\ 1998).
The \ASCA\ spectral fit gives a temperature for the X-ray emitting gas
$kT_{X}\sim0.7\keV$.
Recently the {\sl XMM} RGS spectrum indicates a low temperature component
of $\sim0.6\keV$ (Aschenbach 2002),
and the {\sl XMM} EPIC-PN spectrum indicates $\sim0.9\keV$
(Behar et al.\ 2001).
Compared with the central high temperature Fe~K emission, the low
temperature component could basically ascribed to the shell gas.
If $0.7\keV$ is taken as a mean value of the postshock temperature,
it corresponds to a shell velocity $\vs\sim770\km\ps$,
consistent with the above value derived from the optical observation.

We consider a scenario in which the blast wave propagates rapidly in the
wind cavity until it encounters the cavity wall, where it is
slowed down drastically.
Assuming the swept-up cavity material has been compressed sufficiently,
the volume emission measure, under the thin-shell approximation, is given by
\begin{equation}\label{eq:EM}
EM \approx\frac{(4\no)^{2}}{3}\pi\rs^{3}\lt[1-\lambdj^{3}(1-\beta)\rt],
\end{equation}
where the radius of the remnant $\rs$ is about 12 pc.
For various values of $\beta$ and $\lambdj$, we get from $EM$ the estimates
of the density at the cavity wall, $\no$, and hence the swept-up gas mass,
$\Ms=100M_{\odot}M_{2}$ (eq.[\ref{eq:ma}]), as plotted in Fig.6 together with
other physical parameters obtained from the following calculations.
With the aid of eq.(\ref{eq:fAv1}) for function $\fAv(\lambdj)$, we
then get from eq.(\ref{eq:vAr2}) an estimate of the explosion energy $\Eo$.
The age of the remnant,
$t=10^{3}t_{3}\yr$, consists of two parts, 
$\tj$ (Sedov evolution before the blast wave reaches the cavity wall)
and $t-\tj$ (duration after the impact).
The latter is obtained from eq.(\ref{eq:rAt}) with eq.(\ref{eq:fAr4}) for
function $\fAr(\lambdj)$.

It can be found from Fig.6 that
the kinematic age $t\sim3\E{3}\yr$ and the preshock density
$\no\sim3\cm^{-3}$ can be reproduced by means of the wind cavity model,
if parameters
\[\beta\sim0.1\hspace{5mm}\mbox{and}\hspace{5mm}\lambdj\sim0.94\mbox{-}0.95\]
are adopted.
With this set of parameters, we also have $\Eu\sim3$ and thus avoid an
unreasonably high inference.
As a comparison, if $\beta$ is adopted lower, $\Eu$ would be too high;
and if $\lambdj$ is larger, $t$ would be too large compared with the
kinematic age.
If $\no$ is fixed at $3\cm^{-3}$, the EM (eq.[\ref{eq:EM}]) entails
a density contrast $\beta<0.2$.
We note that, the previous overestimates of $\Eo$ corresponds to the
case that $\fAv(\lambdj)$ is taken as 1 in eq.(\ref{eq:vAr2});
in order for $\Eo$ to be of normal value, $\fAv(\lambdj)$ has to be $>1$.
This condition for $\fAv$ is satisfied around our best-fit parameters
$\beta\sim0.1$ and $\lambdj\sim0.94$-0.95 (see Fig.1),
which implies that the blast wave is still in the course of abrupt
deceleration within the wind-blown shell.

With the best-fit parameters, we also know that it takes the blast wave
$\tj\sim2.2\E{3}\yr$ to travel in the cavity and that 
the velocity of the blast wave just before impacting the cavity wall
is $v_{\rm j}\sim1.9\E{3}\km\ps$ which is decreased to the present value
$\sim800\km\ps$ over a duration of $\sim700\yr$.
Also with these quantities, it is easy to confirm a sufficient compression of
the swept-up cavity material, consistent with the above $EM$ approximation
(eq.[\ref{eq:EM}]).
Moreover, the total swept-up mass is $\Ms\sim170M_{\odot}$, to which the
gas originally inside the cavity contributes $\sim60M_{\odot}$.
The latter mass seems too high for a wind cavity, however this is not
a surprise, if a small part of gas are evaporated from the cavity wall
or/and if interstellar clumps have been left within the cavity so that
the cavity might contain a substantial amount of interstellar material.
The existence of interstellar clumps is reasonable in view of the
proximity of the remnant to the southern molecular cloud (Banas et al.\ 1997).
In fact, a shocked interstellar cloudlet has been found in the remnant
by the {\sl HST} (Blair et al.\ 2000).
In addition, the swept-up mass of the cavity gas ($\sim60M_{\odot}$)
suggests that the SNR has been in the Sedov phase by the time it collides
with the cavity wall.

\section{Conclusion}
Using the thin-shell approximation, we have developed a semi-analytic
model for supernova remnants which evolve crossing a density jump
in the environmental medium.
The generic evolutionary relations are found for two cases,
impact of the blast wave on a cavity wall and breakout of the blast wave
from a dense cloud.
In the impact case, both the adiabatic and radiative phases are investigated;
while in the breakout case, only the adiabatic phase is considered and
the evolution relations are found to be an extension of those in the
former case with different density contrast.
In the impact case, it is also found that the remnant will evolve
rapidly into the radiative phase and even the adiabatic phase
could last so short that it seems to be abortive if the medium density
at the cavity wall is sufficiently high.
The developed model is applied to the cavity-born supernova remnant N132D
whose evolution has not yet been quantitively estimated in a cavity scenario
due to lack of proper model formulae and self-consistent physical parameters
are obtained.

\acknowledgments

We are grateful to Rino Bandiera for helpful advices during this work.
An anonymous referee is thanked for critical comments which have helped
appreciably improve the model application.
John P.\ Hughes is also thanked for helpful comments.
YC and FZ acknowledge support from NSFC grants 1007003 \& 10221001
and grant NKBRSF-G19990754 of China Ministry of Science and Technology.
This work is partially supported by NASA-grant SAO GO-12068X and NASA
LTSA grant NAG5-7935.

\clearpage

\begin{figure}
\plotone{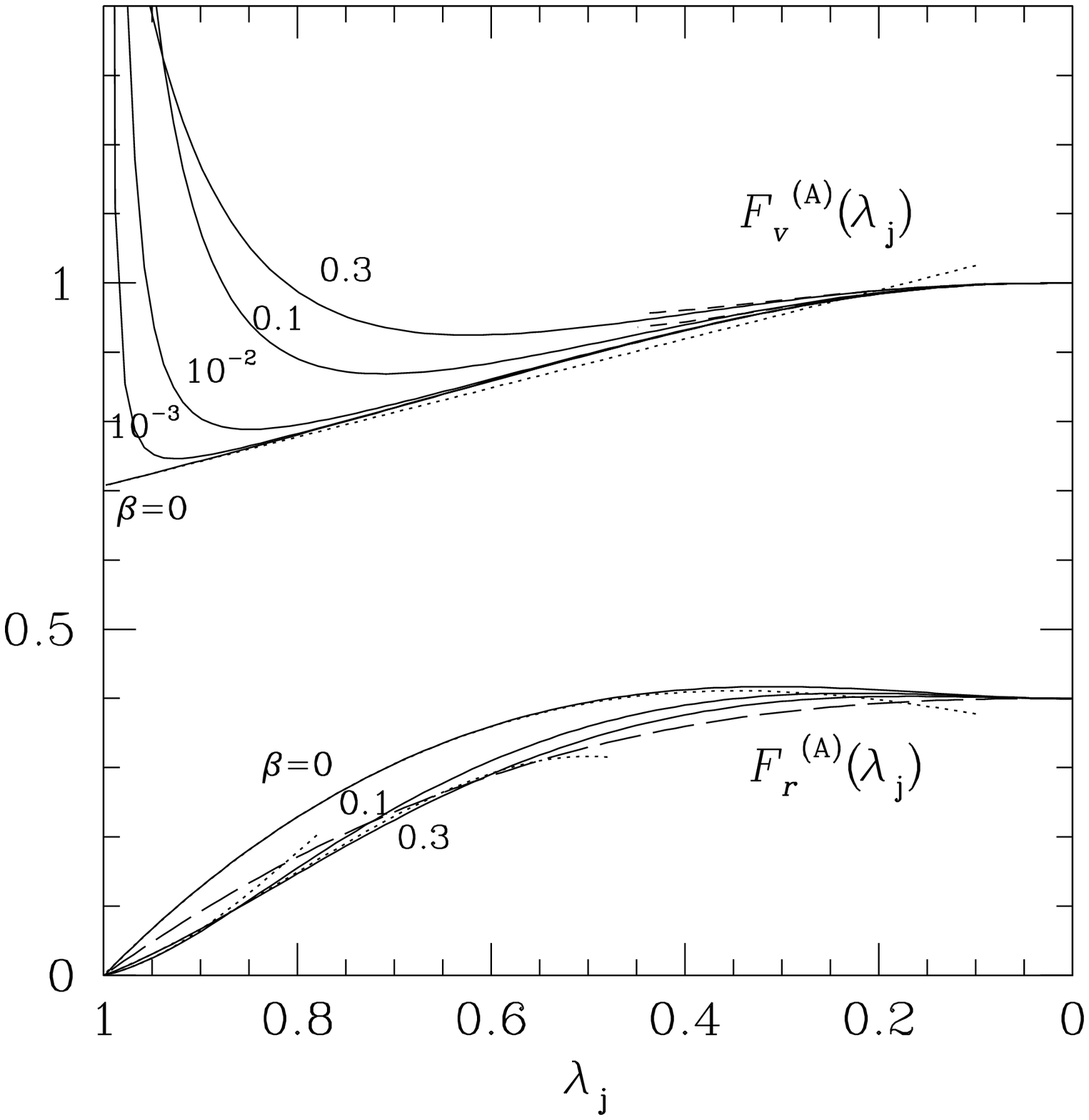}
\caption{Plots of functions $\fAv(\lambdj)$ and $\fAr(\lambdj)$ for
the impact model ($\beta<1$). The solid lines denote the accurate values of
eqs.(\ref{eq:fAv1}) and (\ref{eq:fAr1}).
The four dotted lines, from upper to lower, stand for eq.(\ref{eq:fAv2}) and 
eq.(\ref{eq:fAr2}) in the limit case $\beta=0$, and eq.(\ref{eq:fAr4})
for $\beta=0.1, 0.3$, respectively, while the two short dashed lines
stand for eq.(\ref{eq:fAv3}) in the cases $\beta=0$ and $\beta=0.3$.
The long dashed line is the plot of eq.(\ref{eq:fAr5}).}
\end{figure}
\clearpage

\begin{figure}
\plotone{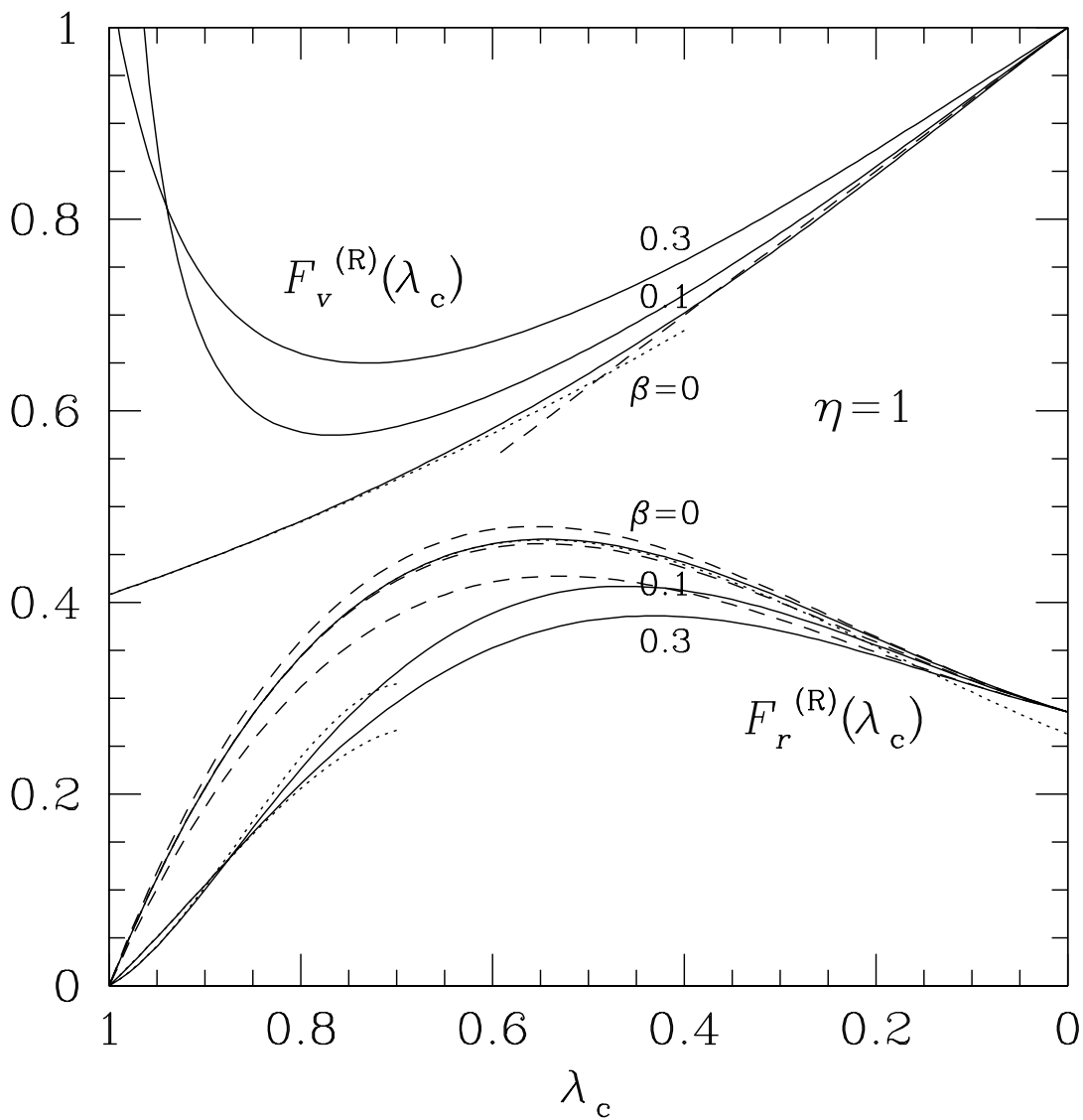}
\caption{Plots of functions $\fRv(\lambdc)$ and
$\fRr(\lambdc)$ for $\eta=1$.
The solid lines denote the accurate values of
eqs.(\ref{eq:fRv1}) and (\ref{eq:fRr1}).
The dotted lines stand for eqs.(\ref{eq:fRv2}) and (\ref{eq:fRr2})
in the case $\beta=0$ and eq.(\ref{eq:fRr4})
in the cases $\beta=0.1,0.3$, respectively.
The dashed lines stand for eq.(\ref{eq:fRv3}) for $\beta=0$
and eq.(\ref{eq:fRr3}) for $\beta=0,0.1,0.3$, respectively.}
\end{figure}
\clearpage

\begin{figure}
\plotone{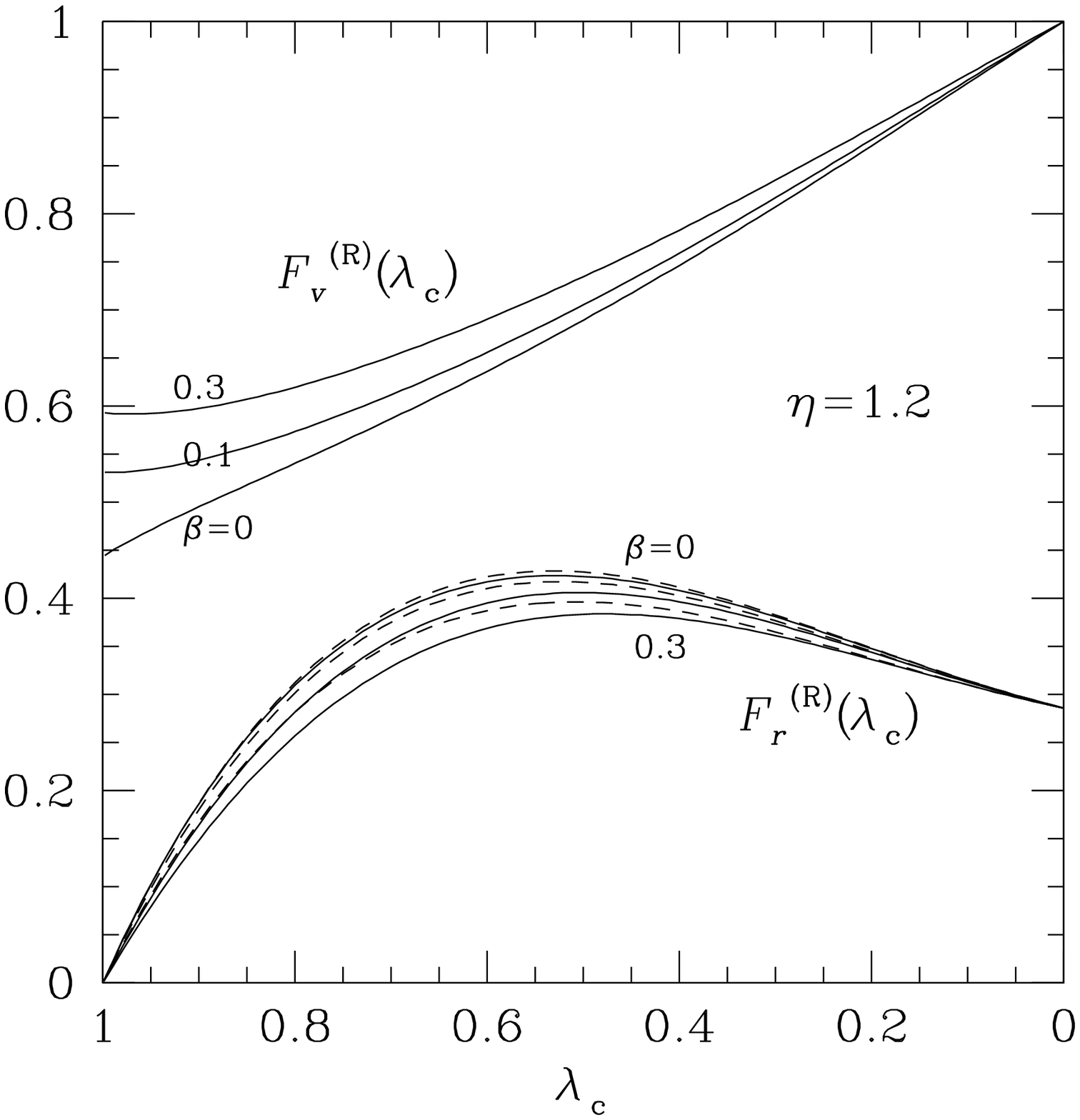}
\caption{Plots of functions $\fRv(\lambdc)$ and
$\fRr(\lambdc)$ for $\eta=1.2$.
The solid lines denote the accurate values of
eqs.(\ref{eq:fRv1}) and (\ref{eq:fRr1}) 
and the dashed lines represent eq.(\ref{eq:fRr3}).
The solid and dashed curves of $\fRr(\lambdc)$ are plotted for
$\beta=0,0.1$, and 0.3, respectively, from upper to lower.}
\end{figure}
\clearpage

\begin{figure}
\plotone{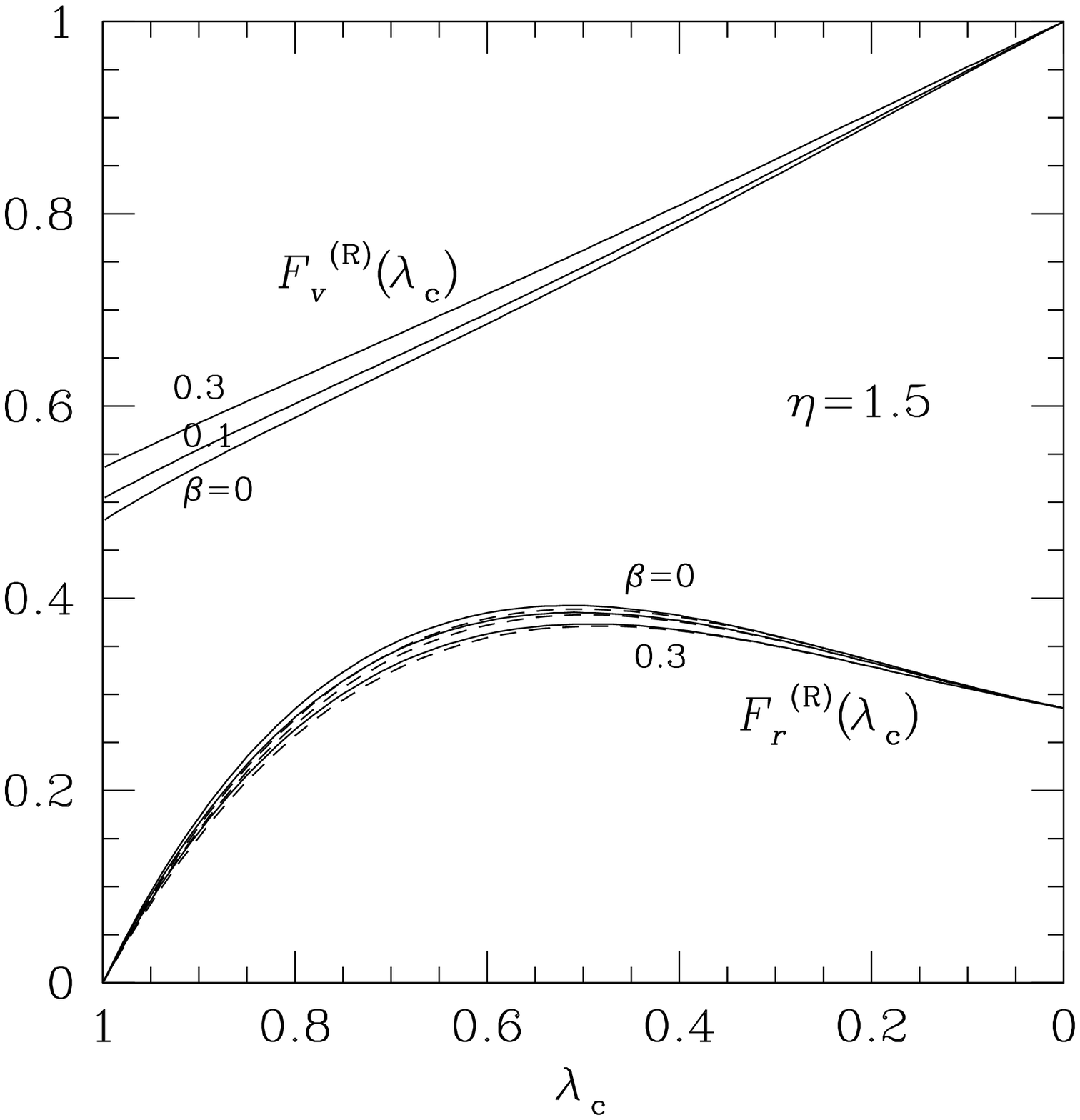}
\caption{Plots of functions $\fRv(\lambdc)$ and
$\fRr(\lambdc)$ for $\eta=1.5$.
The solid lines denote the accurate values of
eqs.(\ref{eq:fRv1}) and (\ref{eq:fRr1}) 
and the dashed lines represent eq.(\ref{eq:fRr3}).
The solid and dashed curves of $\fRr(\lambdc)$ are plotted for
$\beta=0,0.1$, and 0.3, respectively, from upper to lower.}
\end{figure}
\clearpage

\begin{figure}
\plotone{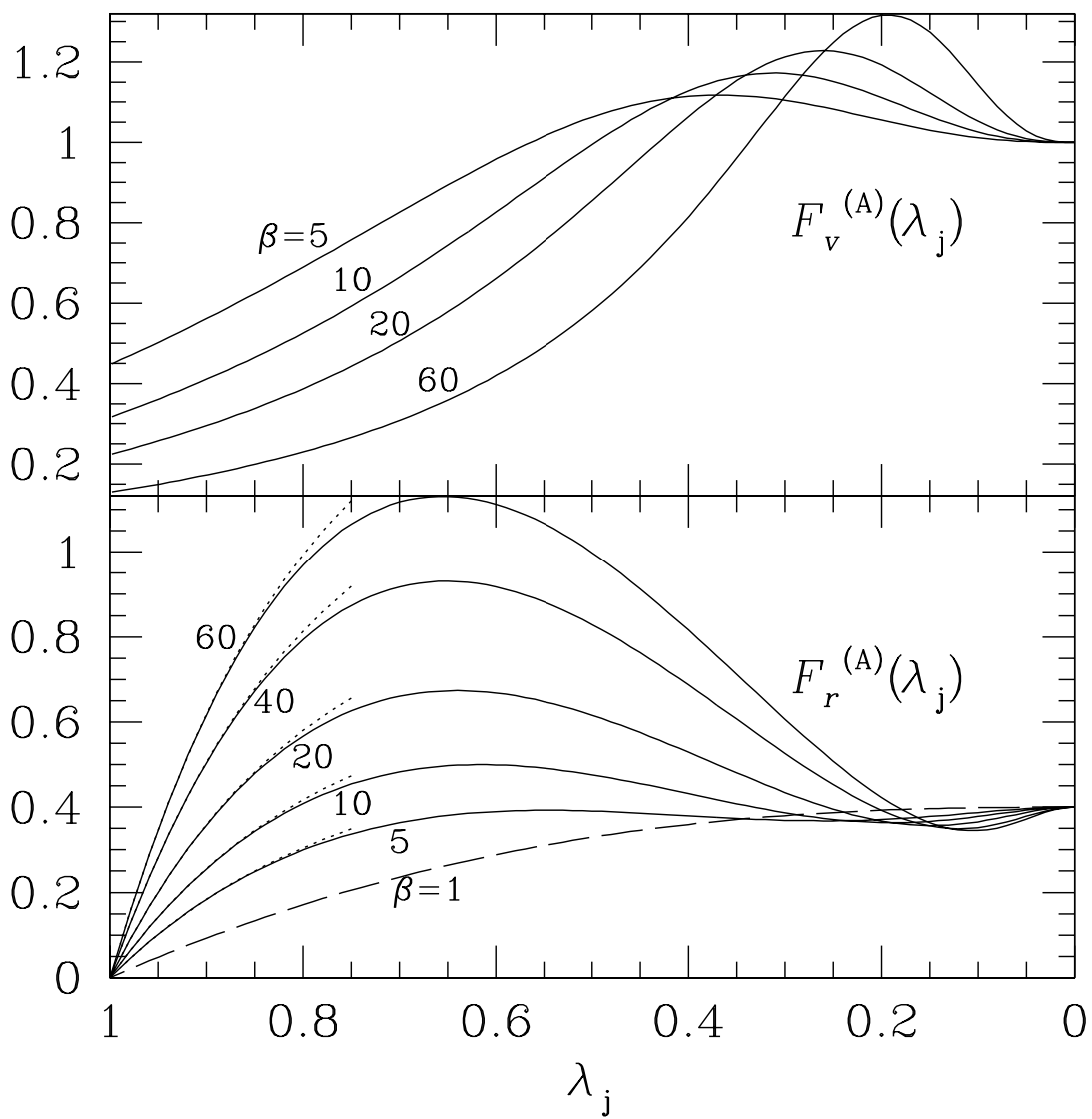}
\caption{Plots of functions $\fRv(\lambdc)$ and
$\fRr(\lambdc)$ for the breakout case ($\beta>1$).
The solid lines denote the accurate values of
eqs.(\ref{eq:fAv1}) and (\ref{eq:fAr1}).
The dotted lines represent the approximate series eq.(\ref{eq:fAr4}).
The long dashed line of $\fAr$ for $\beta=1$ is plotted merely for
comparison.}
\end{figure}
\clearpage

\begin{figure}
\plotone{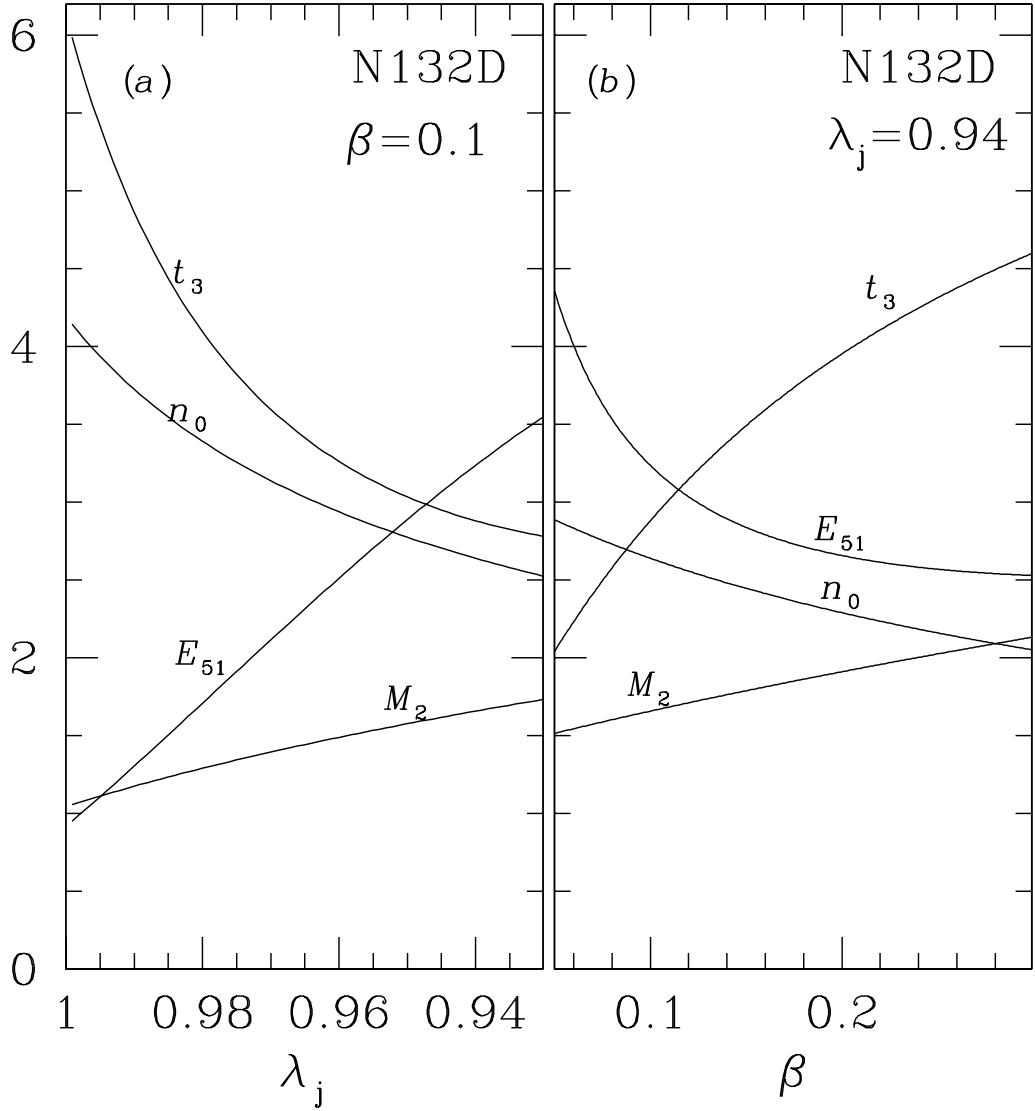}
\caption{Plots of physical parameters derived for SNR~N132D
for $\beta=0.1$ and $\lambdj=0.94$, respectively.}
\end{figure}
\clearpage

\end{document}